\documentclass[reprint,onecolumn,showpacs,superscriptaddress,tighten,nofootinbib,eqsecnum]{revtex4-2}
\usepackage{graphicx,color,hyperref}
\usepackage{dcolumn}
\usepackage{bm}
\usepackage{amsmath}
\usepackage{amsfonts}
\usepackage{here}
\makeatletter

\begin{document}
\title{Proposal for fast computational method for Hertzian contact theory}
\author{Shintaro Hokada}
\affiliation{Department of Industrial Technology and Innovation,
Tokyo University of Agriculture and Technology,
2-24-16, Naka-cho, Koganei, Tokyo 184-8588, Japan}
\author{Shunsuke Iizuka}
\affiliation{Department of Mechanical Systems Engineering, 
Tokyo University of Agriculture and Technology,
2-24-16, Naka-cho, Koganei, Tokyo 184-8588, Japan}
\author{Satoshi Takada}
\email[Corresponding author, e-mail: ]{takada@go.tuat.ac.jp}
\affiliation{Department of Mechanical Systems Engineering, 
Tokyo University of Agriculture and Technology,
2-24-16, Naka-cho, Koganei, Tokyo 184-8588, Japan}
\begin{abstract}
Fast computational method for Hertzian contact theory is proposed.
An incremental formula is introduced to calculate the ellipticity of the contact disk when two elastic bodies are in contact.
This method can determine the ellipticity with good accuracy in a small number of iterations is reported. 
This method is also shown to be applicable from the case of a near perfect circle to the case where the major diameter is sufficiently long compared to the minor diameter.
\end{abstract}

\maketitle
\thispagestyle{empty}
\section{Introduction}
Contact problems between two objects with curvature appear in various fields such as applied physics or engineering such as tribology or powder technology.
Examples in tribology include contact between bearing balls and raceways, or between railroad wheels and rails~\cite{Harris91}.
In powder technology, it is contact between powder particles~\cite{Brilliantov96, Luding08}.

\begin{figure}[H]
    \centering
    \includegraphics[width=0.85\linewidth]{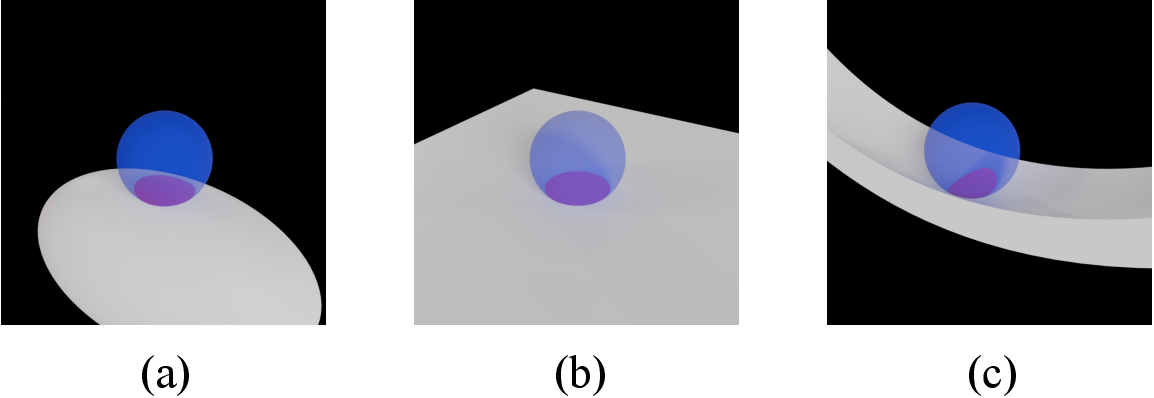}
    \caption{Schematics of contacting two bodies.
    Two bodies have (a) both positive, (b) positive and zero, and (c) positive and negative curvatures.
    Shaded regions represent contact ellipses.}
    \label{fig:contact}
\end{figure}

In general, the contact between two objects with constant curvatures can be divided into three situations depending on the curvatures of the two contacting objects, as shown in Fig.~\ref{fig:contact}.
Hertzian contact theory can calculate the stresses that occur in these situations \cite{Hertz81}.
This theory treats the stress distribution from the degree of deformation of an object by assuming that the forces are balanced at each point in time when contact occurs.

The contact surface predicted by the Hertzian contact theory is generally an ellipse, as can be seen in Fig.~\ref{fig:contact}.
Its shape and stress distribution can be calculated using the complete elliptic integrals.
However, to the best of our knowledge, the ellipticity of such ellipses is not known to be solved analytically.
Therefore, it is generally necessary to rely on numerical calculations.
Various efforts have been made to do so, and indeed, various papers have proposed methods.

The organization of this paper is as follows:
In the next section, we briefly overview the Hertzian contact theory, and introduce previous works relating to this topic.
Section \ref{sec:proposed_method} proposes faster method to determine the shape of contact area, then Sect.~\ref{sec:resutls} validates its efficiency.
In Sect.~\ref{sec:summary}, we summarize our results.

\section{Formulation and previous methods}\label{sec:overview}
This section provides a brief overview of Hertzian contact theory \cite{Hertz81} and an introduction to previous methods describing numerical calculations of the theory \cite{Hamrock73, Hamrock83, Oba14}.
\begin{figure}[htbp]
    \centering
    \includegraphics[width=0.5\linewidth]{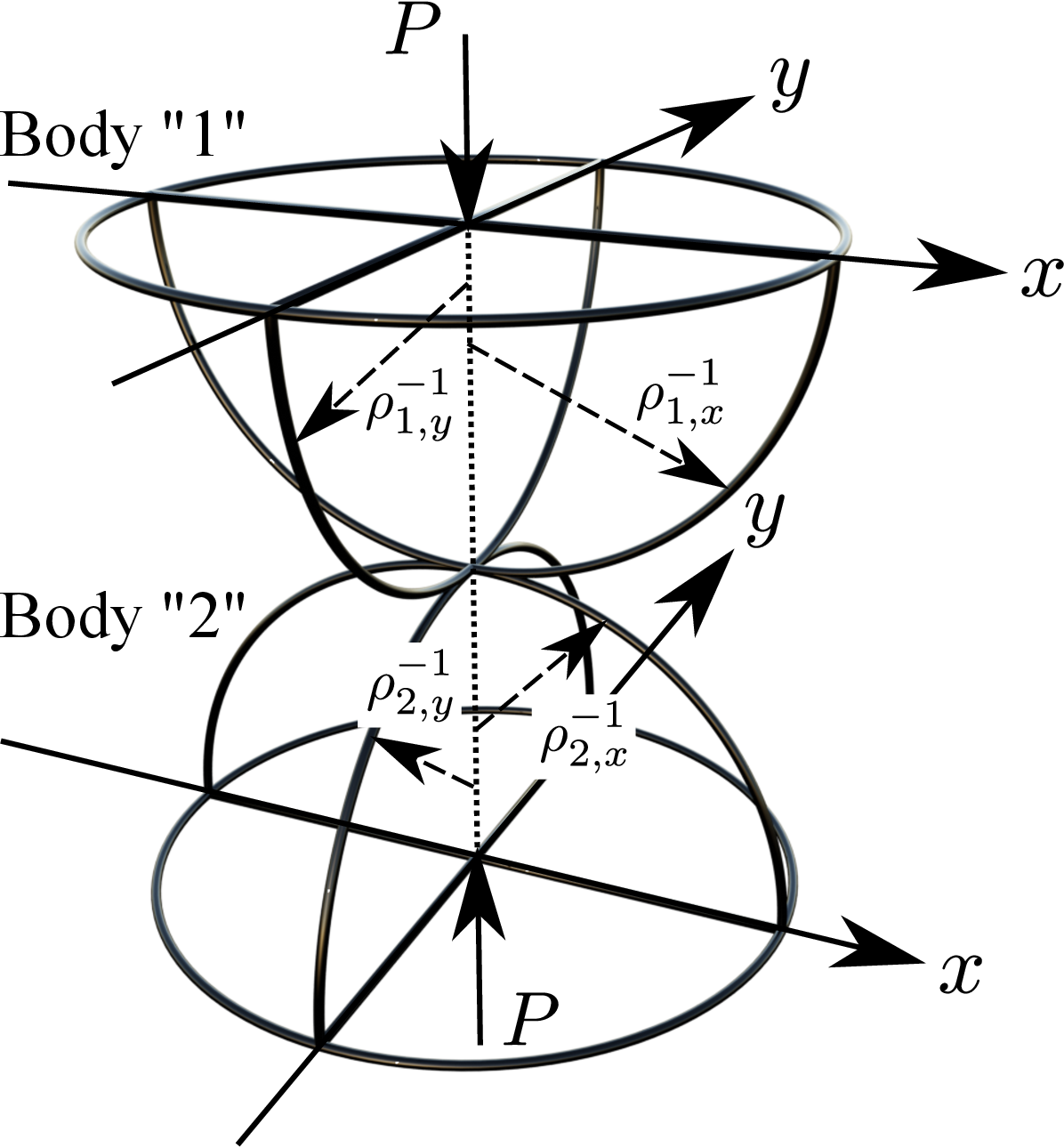}
    \caption{Schematic of two contacting elastic bodies.
    Two bodies are contacted with the force $P$.
    The $x$ and $y$-axes are chosen so that the curvature ratio $\alpha=\rho_x/\rho_y\ge 1$.}
    \label{fig:system}
\end{figure}
As described in the previous section, the contact, as shown in Fig.~\ref{fig:contact}, depends on the curvatures of the two objects.
However, we here generalize them and consider a situation where two objects are in contact as shown in Fig.~\ref{fig:system}.
In this paper, we put the Young's modulus and Poisson's ratio of the two objects as $E_i$ and $\nu_i$ ($i=1,2$), respectively.
Let $\rho_{1,x}$, $\rho_{1,y}$ and $\rho_{2,x}$, $\rho_{2,y}$ be the curvatures of objects 1 and 2 in the $x$ and $y$ axes near the contact point.
In this case, the equivalent curvature $\rho_x$ and $\rho_y$ determined from the two objects are, respectively, defined by
\begin{equation}
    \rho_x = \rho_{1,x}+\rho_{2,x},\quad
    \rho_y = \rho_{1,y}+\rho_{2,y}.
\end{equation}
The $x$-axis and the $x$-axis are written as follows.
However, the $x$ and $y$ axes are defined to be $\rho_x/\rho_y\ge1$.
Using these, the equivalent curvature sum $\rho_+$ and equivalent curvature difference $\rho_-$ can be expressed as
\begin{equation}
    \rho_+\equiv \rho_x + \rho_y,\quad
    \rho_-\equiv \frac{\rho_x-\rho_y}{\rho_+},
\end{equation}
and the equivalent curvature ratio $\alpha$ is introduced by
\begin{equation}
    \alpha \equiv \frac{\rho_x}{\rho_y}
    =\frac{\rho_- + 1}{\rho_- - 1}.
\end{equation}

Let $\ell_x$ and $\ell_y$ be the radius ratios in the $x$ and $y$ axes of the contact ellipses of the two objects in contact, respectively, and we put this ratio as
\begin{equation}
    \lambda \equiv \frac{\ell_y}{\ell_x}.
\end{equation}
In this case, the following relation holds between the radius ratio $\lambda$ and the equivalent curvature ratio $\rho_-$ \cite{Harris91}:
\begin{equation}
    \lambda^2 = \frac{2K(\nu)-E(\nu)(1+\rho_-)}{E(\nu)(1-\rho_-)},
    \label{eq:lambda_rho-}
\end{equation}
with
\begin{equation}
    \nu\equiv 1-\frac{1}{\lambda^2}.
    \label{eq:def_nu}
\end{equation}
Here, $K(\nu)$ and $E(\nu)$ are complete elliptic integrals of the first and second kind, respectively \cite{Abramowitz}
\begin{equation}
    K(\nu)\equiv \int_0^{\pi/2}\frac{\mathrm{d}\theta}{\sqrt{1-\nu \sin^2\theta}},\quad 
    E(\nu)\equiv \int_0^{\pi/2}\sqrt{1-\nu \sin^2\theta}\mathrm{d}\theta.
\end{equation}
Note that $K(\nu)$ is written as $F(\nu)$ and $k=\sqrt{\nu}$ in some literature \cite{Hamrock83, Oba14}, although $\nu$ is a parameter satisfying $0\le \nu\le 1$. 
In this paper, we follow the notation adopted in Mathematica and other software \cite{Abramowitz}.

Usually, the curvatures near the point of contact between two contacting objects are known.
This means that, given $\rho_-$, it is required to calculate the value of $\lambda$ by solving Eq.~\eqref{eq:lambda_rho-}, while it is not easy to solve this equation because $\nu$ appearing on the right side of Eq.~\eqref{eq:lambda_rho-} is a quantity determined by $\lambda$.
However, once $\lambda$ is obtained, $\nu$ is also determined from Eq.~\eqref{eq:def_nu}.
As a result, the contact ellipse radii $\ell_x$ and $\ell_y$ for a given contact load $P$ are
\begin{equation}
    \ell_x = \left(\frac{3E(\nu)P}{\pi \lambda E_\mathrm{eff} \rho_+}\right)^{1/3},\quad
    \ell_y = \left(\frac{3\lambda^2E(\nu)P}{\pi E_\mathrm{eff} \rho_+}\right)^{1/3}
    = \lambda \ell_x,
\end{equation}
where
\begin{equation}
    E_\mathrm{eff}
    \equiv \left(\frac{1-\nu_1^2}{E_1} + \frac{1-\nu_2^2}{E_2}\right)^{-1},
\end{equation} 
is the equivalent Young's modulus determined from two objects.
The surface pressure distribution $\sigma(x,y)$ in the contact ellipse is given by
\begin{equation}
    \sigma(x, y)=\frac{3P}{2\pi \ell_x \ell_y}
    \sqrt{1-\left(\frac{x}{\ell_x}\right)^2-\left(\frac{y}{\ell_y}\right)^2},
\end{equation}
and also for the maximum value of elastic deformation $\delta_\mathrm{max}$ as
\begin{equation}
    \delta_\mathrm{max}
    = \frac{K(\nu)}{2}\left[\frac{9\rho_+}{E(\nu)}
    \left(\frac{P}{\pi\lambda E_\mathrm{eff}}\right)^2\right]^{1/3}.
\end{equation}

The problem here, however, is that the expression \eqref{eq:lambda_rho-} cannot be solved explicitly for $\lambda$, as mentioned in the previous section.
The quantities $\rho_-$ and $\alpha$ determined from the curvatures of the two objects are known quantities, while the ratio of the ellipses at the contact surface $\lambda$ is unknown.
However, the expression \eqref{eq:lambda_rho-} is a formula where $\alpha$ appears when $\lambda$ is determined, and the dependence of $\lambda$ on $\alpha$ cannot be determined in this way.
One way to solve this problem is to transform the formula~\eqref{eq:lambda_rho-}
\begin{equation}
    \alpha = \frac{\lambda^2 E(\nu) - K(\nu)}{K(\nu)-E(\nu)},
    \label{eq:alpha_lambda_relation}
\end{equation}
and substitute various $\lambda$ into it to make a table of $\alpha$ numbers corresponding to each $\lambda$.
This is done by selecting $(\alpha_1, \lambda_1)$ and $(\alpha_2, \lambda_2)$ pairs of $\alpha$ and corresponding $\lambda$ that are close in value in the table, given $\alpha$ for the problem, and then using an appropriate method to calculate $(\alpha, \ lambda)$ in an appropriate way.
However, this method has a disadvantage in that it requires a large amount of computational storage space because it is necessary to calculate values for a large number of $\lambda$ in advance when high accuracy is required.

As a way to compensate for this shortcoming, Hamrock and Anderson \cite{Hamrock73} rewrote Eq.~\eqref{eq:alpha_lambda_relation} as
\begin{equation}
    \lambda = f(\lambda, \alpha),
    \label{eq:zenkashiki}
\end{equation}
and transformed it into the form of an asymptotic equation by letting $\lambda$ on the left side of this equation be $\lambda_{n+1}$ and $\lambda$ on the right side be $\lambda_n$. 
They proposed a method to compute the value by iteratively computing this iteration until the absolute error $|\lambda_{n+1}-\lambda_n|$ or relative error $|\lambda_{n+1}/\lambda_n-1|$ is sufficiently small.    
However, the iterative computation is time-consuming.
To solve this problem, Hamrock and Brewe \cite{Hamrock83} proposed a one-time computation without iterations by carefully choosing the trial function $\lambda(\alpha)$ to be assigned to the right-hand side of Eq.~\eqref{eq:zenkashiki}.
However, even in the range of $1\le \alpha\le 10$, which is a subject of wide interest, it contains an error of about $10~\%$, which is not satisfactory for calculations that require accuracy.

In contrast, Oba \cite{Oba14} proposed an improved version of the trial function proposed by Hamrock and Anderson \cite{Hamrock73} and used it to obtain $\alpha$ from Eq.~\eqref{eq:alpha_lambda_relation}, and substituting it into the trial function.
His method achieves an accuracy of about $10^{-10}$ with a small number of iterations (3 iterations), and is faster than conventional methods.

\section{Proposed method}\label{sec:proposed_method}
We improve on Oba's work \cite{Oba14} and propose a method that produces the same level of performance with fewer iterations.
First, let us discuss the computation of elliptic integrals which appears in Eq.~\eqref{eq:alpha_lambda_relation}.
Some previous literature \cite{Hamrock83, Oba14} introduced the fast computational method of elliptic integrals.
However, most programming languages have libraries that can compute elliptic integrals, and it is expected to be faster than writing the code by oneself.
Therefore, in this paper, we do not discuss this issue, but assume that a suitable library is available to compute elliptic integrals.
However, if one's environment requires the computation of elliptic integrals, one can use Hastings formulas \cite{Fukushima09, Oba14} or other formulas \cite{Whittaker} with appropriate degree termination.

\begin{figure}[htbp]
    \centering
    \includegraphics[width=0.75\linewidth]{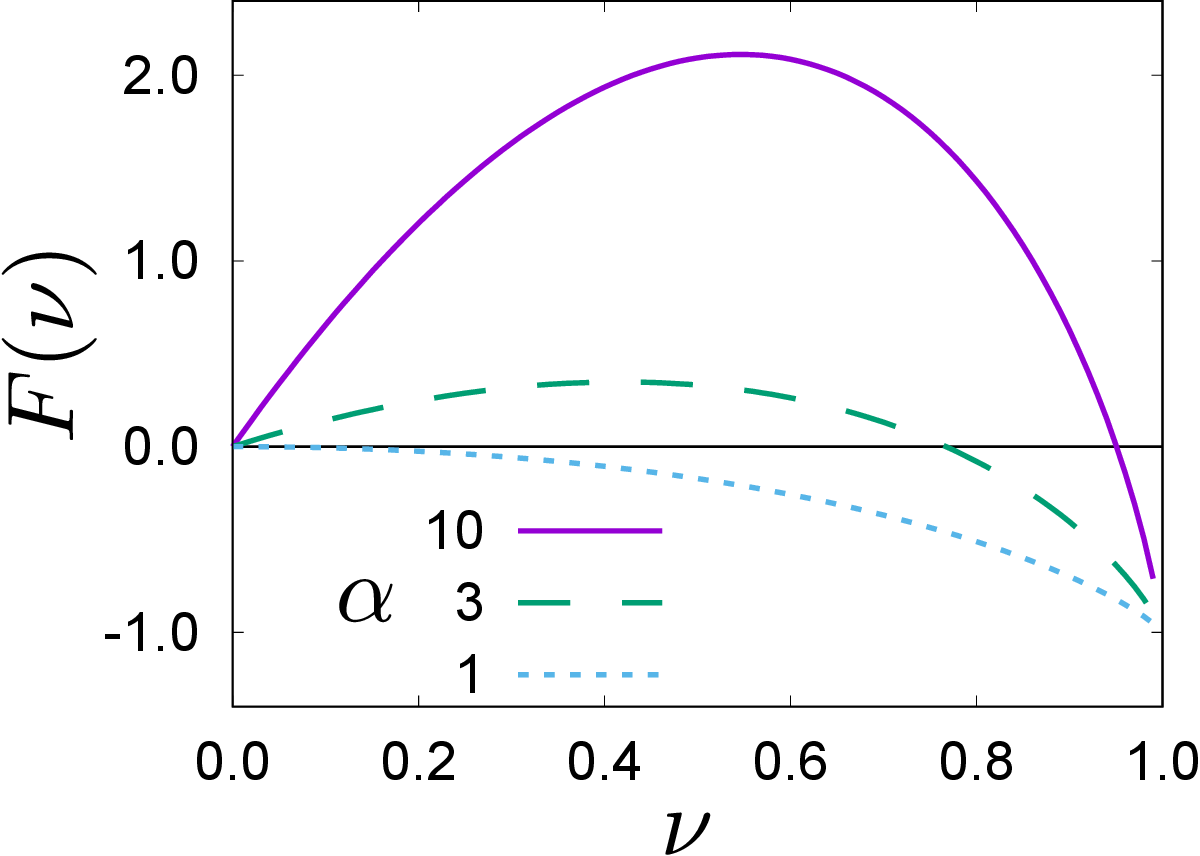}
    \caption{Plots of $F(\nu)$ against $\nu$ for $\alpha=1$ (dotted line), $3$ (dashed line), and $10$ (solid line).}
    \label{fig:F_nu}
\end{figure}

We first describe how to determine $\nu$ that satisfies Eq.~\eqref{eq:alpha_lambda_relation} for $\alpha$.
This is because $\lambda$ is also determined when $\nu$ is determined from Eq.~\eqref{eq:def_nu}.
Using Eq.~\eqref{eq:def_nu}, Eq.~\eqref{eq:alpha_lambda_relation} can be expressed as
\begin{equation}
    F(\nu)
    \equiv (\alpha+1)(1-\nu)K(\nu) 
    - [1+\alpha(1-\nu)]E(\nu)=0.
\end{equation}
Then $\nu$ is a solution satisfying $0\le \nu\le 1$ with $F(\nu)=0$.
Figure \ref{fig:F_nu} shows the graphs of $F(\nu)$ for $\alpha=1$, $3$, and $10$.
It can be seen that $\alpha=1$ is a monotonically decreasing function and only $\nu=0$ satisfies $F(\nu)=0$, whereas $\nu=0$ and $0<\nu< 1$ exist for $\alpha>1$.
However, since $\nu=0$ is the case where the elliptic integrals and the corresponding elliptic functions are attributed to trigonometric functions, the appropriate solution for $\alpha>1$ is the one satisfying $0<\nu \le 1$.
This solution approaches $\nu\to1$ as $\alpha$ increases.
In other words, $\lambda$ increases from Eq.~\eqref{eq:def_nu}.

The solution is obtained by an iterative method.
Here, we adopt Bailey's method, which converges more quickly than Newton's method \cite{Traub61}.
First, an appropriate $\nu_0$ is determined.
Next, given $\nu_n$, we can get $\nu_{n+1}$ by the Bailey method
\begin{equation}
    \nu_{n+1}
    = \nu_n - \frac{F(\nu_n)}{F^\prime(\nu_n)-\dfrac{F(\nu_n)F^{\prime\prime}(\nu_n)}{2F^\prime(\nu_n)}},
    \label{eq:recurrence}
\end{equation}
where
\begin{subequations}
\begin{align}
    F^\prime(\nu)
    &= \frac{-(1+2\alpha)K(\nu)+3\alpha E(\nu)}{2},\\
    F^{\prime\prime}(\nu)
    &= -\frac{(1-\nu)(\alpha-1)K(\nu)+[1-\alpha(1-3\nu)]E(\nu)}{4\nu(1-\nu)}.
\end{align}
\end{subequations}
In Bailey's method, it is necessary to start the calculation from a good initial condition $\nu_0$, which is somewhat close to a true value.
This is because if $\alpha>1$, starting from $\nu_0$ in the region $F^\prime(\nu)>0$, the calculation will converge to $\nu=0$.
In order to speed up the convergence, it is necessary to start as close as possible to the true solution.

For the above purposes, we set up an approximate formula for the relationship between $\alpha$ and $\nu$ and start the Bailey method with that value.
However, when we plot the relationship between $\alpha$ and $\nu$ from Eq.~\eqref{eq:alpha_lambda_relation}, the change is large in $\nu\simeq 0$ and $1$.
Therefore, we use an equation to approximate the $\alpha$ dependence of $\lambda$.
In this study, we adopt the trial function proposed by Ref.~\cite{Oba14}:
\begin{equation}
    \log \lambda_0 
    = \log \alpha \cdot 
    \left[c_1 + c_2 (\log \alpha)^{1/2}
    + c_3 (\log \alpha)^{3/4}
    + c_4 (\log \alpha)\right],
    \label{eq:Oba_fitting}
\end{equation}
where the coefficients
\begin{equation}
    c_1=0.666947681,\ 
    c_2=0.082186314,\ 
    c_3=0.109262836,\ 
    c_4=0.028024325,
\end{equation}
are the same as those given in Ref.~\cite{Oba14}.

Summarizing the above, we propose the following procedure for the actual computation
\begin{enumerate}
    \item Calculate $\lambda_0$ from Eq.~\eqref{eq:Oba_fitting} for $\alpha$ to be calculated and obtain $\nu_0=1-1/\lambda_0^2$.
    \item Calculate $\nu_{n+1}$ by substituting $\nu_n$ ($n=0,1,\cdots$) into Eq.~\eqref{eq:recurrence}.
    \item Repeat (ii), until $|\nu_{n+1}-\nu_n|<\delta_\mathrm{th}$ is satisfied for a certain threshold $\delta_\mathrm{th}$.
\end{enumerate}

\section{Results}\label{sec:resutls}
\begin{figure}[htbp]
    \centering
    \includegraphics[width=0.75\linewidth]{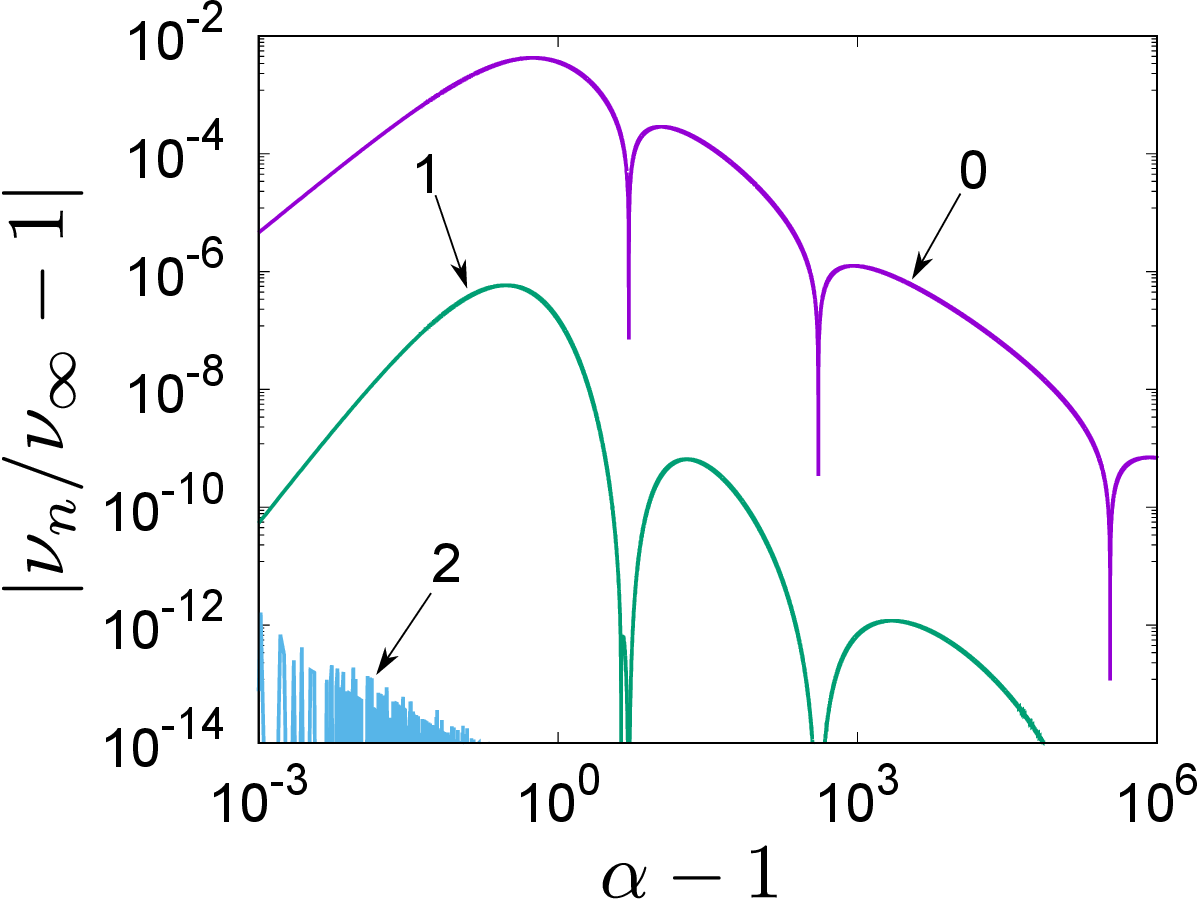}
    \caption{Convergence of $\nu_n$ to $\nu_\infty$ against $\alpha$ when the number of iterations is $0$, $1$, and $2$.}
    \label{fig:convergence}
\end{figure}

This section presents the main results obtained using the method described in the previous section.
In order to see the convergence of the results, we regard $\nu_n$ as a true value when $|\nu_{n+1}-\nu_n|<1.0\times 10^{-15}$ is satisfied, and denote it as $\nu_\infty$.
Figure \ref{fig:convergence} plots the relative error $|\nu_n/\nu_\infty-1|$ from the true value for each iteration $n$ for each $\alpha$.

The results show that the error is less than $10^{-12}$ in two iterations for $10^{-3}\le \alpha \le 10^6$.
If the error tolerance can be relaxed to $10^{-6}$, only one interation is needed.
The previous paper \cite{Oba14} reported that when the parameter $\nu$ is close to $0$, which means that the contact ellipse becomes an almost perfect circle, the convergence is deteriorated due to the drop of digits caused by the small values of the denominator and numerator of the asymptotic equation.
In this case, a different formula that is a variant of the asymptotic formula was used.
In contrast, the new method does not have such a case separation.
This is another advantage of this method.

\section{Conclusion}\label{sec:summary}
In this paper, we discuss a fast method for computing the elliptic integrals that appear when calculating the shape of the contact ellipse and the stress distribution at the contact between two elastic bodies.
The method of Oba \cite{Oba14} is modified, and the iteration equation is given for the parameter $\nu$ of the ellipse.
By solving it with Bailey's method, we proposed a method in which the error relative to the exact value is less than $10^{-12}$ in two iterations.
When the threshold is enough to be $10^{-6}$, this method is sufficient for a single calculation, and it is also practical for calculations at each time point.

\section*{Acknowledgement}
The authors thank Yasuhisa Ando for his fruitful suggestion.
This work is supported by the Grant-in-Aid of MEXT for Scientific Research (Grant No.~\href{https://kaken.nii.ac.jp/en/grant/KAKENHI-PROJECT-24K06974/}{24K06974} and No.~\href{https://kaken.nii.ac.jp/en/grant/KAKENHI-PROJECT-24K07193/}{24K07193}).



\begin{thebibliography}{99}
\bibitem{Harris91}
    T. A. Harris,
    {\it Rolling Bearing Analysis Third Edition}
    (Wiley, New York, 1991).
\bibitem{Brilliantov96}
    N. V. Brilliantov, F. Spahn, J.-M. Hertzsch, and T. P\"{o}schel,
    \href{https://doi.org/10.1103/PhysRevE.53.5382}
    {Phys. Rev. E \textbf{53}, 5382 (1996)}.
\bibitem{Luding08}
    S. Luding,
    \href{https://doi.org/10.1007/s10035-008-0099-x}
    {Granul. Matter \textbf{10}, 235 (2008)}. 
\bibitem{Hertz81}
    H. Hertz,
    \href{https://doi.org/10.1515/crll.1882.92.156}
    {J. Reine Angew. Math. \textbf{92}, 156 (1881)}.

\bibitem{Hamrock73}
    B. J. Hamrock and W. J. Anderson, 
    \href{https://doi.org/10.1115/1.3451796}
    {J. Lubr. Technol. \textbf{95}, 265 (1973)}. 

\bibitem{Hamrock83}
    B. J. Hamrock and D. Brewe,
    \href{https://doi.org/10.1115/1.3254558}
    {J. Lubr. Technol. \textbf{105}, 171 (1983)}. 
\bibitem{Oba14}
    M. Oba, 
    \href{https://doi.org/10.1299/transjsme.2014cm0105}
    {Trans. JSME \textbf{80}, CM0105 (2014)}.

\bibitem{Abramowitz}
    M. Abramowitz and I. A. Stegun, 
    {\it Handbook of Mathematical Functions: With Formulas, Graphs, and Mathematical Tables}
    (Dover, 1964).
\bibitem{Fukushima09}
    T. Fukushima,
    \href{https://doi.org/10.1007/s10569-009-9228-z}
    {Celest. Mech. Dyn. Astr. \textbf{105}, 305 (2009)}. 
\bibitem{Whittaker}
    E. T. Whittaker and G. N. Watson,
    {\it A Course of Modern Analysis}
    (Cambridge Univ. Press, Cambridge, 1927).
\bibitem{Traub61}
    J. F. Traub,
    \href{https://doi.org/10.1145/366199.366252}
    {Commun. ACM \textbf{4}, 143 (1961)}.
\end{thebibliography}
\end{document}